# Micro-Pattern Gas Detectors for Charged-Particle Tracking and Muon Detection

M. Hohlmann[1], V. Polychronakos[2], A. White[3], and J. Yu[3]

[1]Dept. of Physics and Space Sciences, Florida Institute of Technology, Melbourne, FL 32901
[2]Brookhaven National Laboratory, Upton, NY 11973
[3]Dept. of Physics, University of Texas, Arlington, TX 76019

June 8, 2013

*Abstract* – **In the context of the 2013 APS-DPF Snowmass summer study conducted by the U.S. HEP community, this white paper outlines a roadmap for further development of Micro-pattern Gas Detectors for tracking and muon detection in HEP experiments. We briefly discuss technical requirements and summarize current capabilities of these detectors with a focus of operation in experiments at the energy frontier in the medium-term to long-term future. Some key directions for future R&D on Micro-pattern Gas Detectors in the U.S. are suggested.**

## I. Physics drivers and justification

Exploration of virtually all physics at the energy frontier needs high-performing particle tracking – even jet physics since jet reconstruction in calorimeters with particle flow algorithms has now become standard. Study of any physics process with at least one high-$p_T$ muon in the final state requires robust muon identification and triggering. These capabilities are core requirements for all physics at the current and future energy frontiers, such as searches for new physics and standard model precision measurements, in particular in the newly opened Higgs sector.

The high-rate environments at the current energy frontier, i.e. LHC phases 1 and 2, and at future high-energy colliders (high-energy LHC, ILC, CLIC, muon collider) drive much of the development of advanced Micro-pattern Gas Detectors (MPGD's) to provide robust particle tracking, muon identification and muon triggering in increasingly harsh radiation environments. We note that applications of MPGD's also extend into radiation detection and particle tracking in areas outside of HEP and NP, e.g. medical imaging [1] and homeland security [2].

## II. Technical requirements on advanced MPGD's for tracking and muon detection

Muon detectors in HEP experiments typically cover areas of many square meters. Consequently, economic construction of large-area MPGD's is mandatory for muon detector systems. The anode structures for signal pickup in MPGDs should be optimized to save cost by minimizing the required number of electronics channels while maintaining or improving performance. Using MPGD's as tracking detectors requires highest tolerance against radiation damage and high-rate capability. We suggest a rate capability of $\mathcal{O}(100 \text{ MHz/cm}^2)$ at minimal discharge rates as a benchmark R&D goal. Lowest detector mass is desired to minimize multiple scattering and bremsstrahlung in trackers. Very high spatial resolution of $\mathcal{O}(10 \text{ μm})$ for normal incidence could make MPGD trackers competitive with silicon-based trackers in terms of performance, but at potentially considerably lower cost. High detection efficiencies for charged particles very near to 100% and good time resolution of $\mathcal{O}(1 \text{ ns})$ will enable fully efficient track and muon triggering.

*Table 1: Typical current MPGD capabilities and suggested design and performance goals*

|  | Current capability | Suggested design/performance goal |
|---|---|---|
| **Spatial resolution** (normal incidence) | ~ 50 μm | ~ 10 μm |
| **Timing resolution** | ~ 3 ns | 1 ns |
| **Rate capability** | 12 MHz/cm$^2$ | 100 MHz/cm$^2$ |
| **Module size** | 0.5-1 m$^2$ | > 2 m$^2$ |
| **Number of readout channels** | Typically 12-25 per cm (400-800 μm strip pitch) | ~ 5 per cm |
| **Module cost** | ~$10k per m$^2$ | ~$1k per m$^2$ |

## III. Current technical capabilities

MPGD's such as Gas Electron Multipliers (GEMs) [3] and Micromegas [4] have been operating for years with good stability in HEP experiments such as COMPASS [5,6], TOTEM [7,8], and LHCb [9,10]. GEMs have run at muon rates up to 12 MHz/cm$^2$ in COMPASS [11]. Best resolutions achieved with regular anode strip readouts are around 50 μm (COMPASS) [12] for normally incident charged particles, and 3 ns (LHCb) [10]. Higher spatial resolutions of ~ 25 μm have been achieved for small areas of a few cm$^2$ by coupling MPGDs directly to CMOS pixel chips [13]. MPGD's allow detectors to be configured in planar, cylindrical [14], or spherical [15] geometries. Chambers are thin with heights typically 1 cm or less. Anode patterns take many forms from simple microstrips to complex two-dimensional patterns. MPGDs are now reaching lengths of 1-2 m at the prototype level for ATLAS and CMS [16-18]. Tab. 1 contrasts typical current MPGD capabilities with the performance goals outlined in the previous section. Goals depend on the specific application and do not necessarily have to be all achieved simultaneously.

## IV. Key R&D directions

We suggest R&D in the following key directions to achieve the design and performance goals for MPGDs and to make them cost-effective alternatives to silicon vertex and tracking detectors:

- Development of
  - innovative signal induction structures (anodes)
  - resistive signal induction structures for highest stability with respect to discharges
  - highly integrated rad-hard frontend readout electronics with at least 4k channels/chip
  - detector-electronics interconnects with highest densities
  - multi-GHz sampling of induced signal charges with multi-channel readouts to resolve ionization clusters in time and space for improving temporal and spatial resolutions.
- Integration of flex-circuit readout electronics into the MPGD structures.
- Studies of material selection and aging behavior for highest radiation loads.
- Development of cost-effective MPGD construction techniques for detector mass production.
- Investigation of alternative production techniques such as additive manufacturing for detection elements and possibly entire detectors.
- Development of an industry base with MPGD mass production capabilities within the US.

## V. Conclusion

Micro-pattern gas detector technology has been proven for tracking charged particles and for detecting and triggering on muons in high-rate environments. Large-area MPGD's are now becoming available. This makes MPGD's an attractive option for tracking and muon instrumentation in future experiments at the energy frontier. With further development, MPGD's could have the potential for performing as well as silicon detectors in tracking and vertexing applications, but at lower cost. Consequently, we recommend a strengthening of R&D efforts geared towards advancing MPGD's within the U.S. HEP community.

## References

[1] R. M. Gutierrez, et al., "MPGD for breast cancer prevention: A high resolution and low dose radiation medical imaging," J. Instrum., 7 (2012).

[2] K. Gnanvo, et al., "Imaging of high-Z material for nuclear contraband detection with a minimal prototype of a muon tomography station based on GEM detectors," Nucl. Instr. Meth. A, 652 (2011) 16-20.

[3] F. Sauli, "GEM: A new concept for electron amplification in gas detectors," Nucl. Instr. Meth. A, 386 (1997) 531-534.

[4] Y. Giomataris, et al., "MICROMEGAS: A high-granularity position-sensitive gaseous detector for high particle-flux environments," Nucl. Instr. Meth. A, 376 (1996) 29-35.

[5] P. Abbon, et al., "The COMPASS experiment at CERN," Nucl. Instr. Meth. A, 577 (2007) 455-518.

[6] A. Austregesilo, et al.,"First Results of the PixelGEM Central Tracking System for COMPASS," Nucl. Phys. B Proc. Suppl. 197 (2009) 113-116.

[7] J. Baechler, et al., "The TOTEM experiment at LHC," Proc. of 2011 IEEE Nuclear Science Symposium and Medical Imaging Conference, Valencia, Spain, Oct. 23-29, 2011, p. 1417-1420.

[8] M. Quinto, et al., "The TOTEM GEM Telescope (T2) at the LHC," Nucl. Phys. B Proc. Suppl., 215 (2011) 225-227.

[9] A. Augusto Alves Jr., et al.,"The LHCb detector at the LHC," J. Instrum. 3 (2008).

[10] A. Cardini, et al., "The Operational Experience of the Triple-GEM Detectors of the LHCb Muon System: Summary of 2 Years of Data Taking," Proc. of 2012 IEEE Nuclear Science Symposium and Medical Imaging Conference, Anaheim, CA, Oct. 29 – Nov. 3, 2012, p. 759.

[11] B. Ketzer, "Status of PixelGEM Detectors," COMPASS Technical Board Meeting (2011, unpublished).

[12] B. Ketzer, et al.,"A fast tracker for COMPASS based on the GEM," Nucl. Phys. B Proc. Suppl. 125 (2003) 368-373.

[13] M. Titov, "New developments and future perspectives of gaseous detectors," Nucl. Instr. Meth. A, 581 (2007) 25-37.

[14] A. Balla, et al., "Design and construction of a cylindrical GEM detector as Inner Tracker in KLOE-2," Proc. of 2011 IEEE Nuclear Science Symposium and Medical Imaging Conference, Valencia, Spain, Oct. 23-29, 2011, p. 1002-1005.

[15] S. Duarte Pinto, et al., "Making spherical GEMs," J. Instrum., 4 (2009).

[16] J. Wotschack, "The development of large-area micromegas detectors for the ATLAS upgrade," Mod. Phys. Lett. A, (2013) 1-11.

[17] D. Abbaneo, et al., "An overview of the design, construction and performance of large area triple-GEM prototypes for future upgrades of the CMS forward muon system," J. Instrum., 7 (2012).

[18] D. Abbaneo, et al., "Beam Test Results for New Full-scale GEM Prototypes for a Future Upgrade of the CMS High-eta Muon System," Proc. of 2012 IEEE Nuclear Science Symposium, Anaheim, CA, Oct. 29 - Nov. 3, p. 1172.